\documentclass[pra,aps,showpacs,superscriptaddress]{revtex4}
\usepackage{amsmath}
\usepackage{amsfonts}
\usepackage{amssymb}
\usepackage{mathrsfs}
\usepackage{epsfig}
\usepackage{bm}
\usepackage{amsbsy}
\usepackage{color}
\usepackage{epstopdf}
\usepackage{txfonts}

\newcommand{\ket}[1]{\vert#1\rangle}
\newcommand{\bra}[1]{\langle#1\vert}

\newcommand{\ketbra}[2]{\vert#1\rangle\langle#2\vert}

\begin{document}


\title{Quantum circuits for spin and flavor degrees of freedom of quarks forming nucleons}

\author{Fernando L. Semi\~ao}
\affiliation{{Departamento de F\'isica, Universidade Estadual de Ponta Grossa, Campus Uvaranas, 84030-900, Ponta Grossa, PR, Brazil}}
\author{Mauro Paternostro}
\affiliation{School of Mathematics and Physics, The Queen's University, Belfast,
BT7 1NN, UK}
\date{\today}

\begin{abstract}
We discuss the quantum-circuit realization of the state of a nucleon in the scope of simple simmetry groups. Explicit algorithms are presented for the preparation of the state of a neutron or a proton as resulting from the composition of their quark constituents. We estimate the computational resources required for such a simulation and design a photonic network for its implementation. Moreover, we highlight that current work on three-body interactions in lattices of interacting qubits, combined with the measurement-based paradigm for quantum information processing, may also be suitable for the implementation of these nucleonic spin states.
\end{abstract}


\maketitle

The last decade has seen a fervent activity in the {\it simulation} of complex quantum phenomena through simple and fully controllable systems. Examples of quantum simulations are now aplenty: the superfluid-Mott insulator quantum phase transition can be simulated using neutral atoms loading optical lattices~\cite{mott} or arrays of coupled cavities with embedded two-level systems~\cite{cc}. Molecular energies have been efficiently computed using quantum algorithms and those of $\text{H}_2$  reproduced in a photonic quantum simulator~\cite{emol}. Gravitational black holes, Hawking radiation and the Unruh effect have found fertile ground for their simulation in Bose-Einstein condensates~\cite{bec}. Atoms interacting in special ways with light have been used to theoretically simulate important models in condensed-matter theory, such as the Lieb-Liniger one~\cite{morigi}, while Dirac's predictions on the {\it Zitterbewegung} have been reproduced in various contexts~\cite{zitter,dirac}.

 The interest in quantum simulation is manyfold. First, important information on the statistical behavior and properties of the simulated mechanisms is gathered by dealing with systems that are more easily manipulated and measured. In this sense, the development of quantum computing is gaining further significance as a valuable tool that sheds light on difficult problems studied from a new (information-theoretical) viewpoint. Second, we are going towards the actual realization of the original idea of quantum simulations put forward by Feynman~\cite{feynman}. Yet, this very same concept has considerably evolved since then. Deutsch has recently pointed out that  \emph{``[quantum simulation] would be used for smaller things, not things on a larger scale than a molecule...Small molecules and interactions within an atom, subtle differences between different isotopes [...]. And of course things on an even smaller scale than that''}~\cite{deutsch}. Buluta and Nori remarked that quan
 tum simulators would not only provide new results (hardly achievable classically), but also allow to test models whose experimental access is either too expensive or beyond the reach of current technology, like {string theory}~\cite{buluta}. 

In this paper we move along such lines and propose a simple protocol for mimicking the $\text{SU}(4)$ quark model of nucleonic spin states based on the {\it digital quantum simulator} approach~\cite{buluta}. We use qubits to encode the information carried by quarks and give a quantum-circuit version of the algorithm needed for the achievement of the spin-up protonic and neutronic states as described in Refs.~\cite{qlbook}. Our proposal is economic in terms of resources required and very flexible: we design a photonic network based on the use of only six modes, which is well within current experimental capabilities, where the spin-up state of a proton or a neutron could be simulated. Moreover, we discuss how an observable emulating the intrinsic nucleonic magnetic moment can be measured and we highlight the possibility for implementation in an optical lattice-based quantum simulator. It should be emphasized that bare nucleonic states (without including gluons), typical of the first quark models, are not the whole picture. Quantum chromodynamics (QCD) is clearly the most rigorous theoretical apparatus for the description of nucleons.  In spite of their inherent limitations, the use of simple quark models has often been very successful in capturing the \emph{general} nucleonic properties. In fact, these models allow for a very reasonable understanding of many aspects of their spectroscopy ~\cite{old}, and a possible theoretical explanation for such agreement is presented in \cite{morpurgo}. Our goal here is to make a first step towards the simulation of subatomic particles, starting with the spin state of neutrons and protons in the scope these simplified models.


\emph{Quark Model for Protons and Neutrons}. - In 1961, Ne'eman and, independently, Gell-Mann and Nishijima introduced a classification scheme for light hadrons, now known as the Eightfold way, based on the use of their charge $Q$ and strangeness $S_{\!t}$.  Charge and strangeness are examples of quantum numbers associated to internal symmetries, i.e. symmetries following from transformations that do not involve space and time. Other examples are the baryon number $B$ and the isospin $I$. All such quantum numbers are necessary for the understanding of subatomic reactions driven by strong and weak forces. Not all the internal quantum numbers used to describe baryons in the Eightfold way are independent for they are bound to the empirical Gell-Mann-Nishijima equation $Q=I_3+\tfrac{1}{2}(B+S_{\!t})$, where $I_3$ is the projection of the isospin on an arbitrary axis.  

In line with the Eightfold way, hadrons can be divided in two groups, according to their spin: \emph{baryons} (fermions) and \emph{mesons} (bosons). Gell-Mann-Nishijima formula can be applied
equally to baryons and mesons, thus suggesting the existence of a fundamental explanation of the symmetries in the Eightfold way and the range of values that each quantum number can take. Such features can indeed be explained by assuming that hadrons are formed by the \emph{combination} of a small number of constituents called quarks. In the first quark model~\cite{su3}, three types of quarks were proposed to accommodate the known baryons and mesons. They are called \emph{flavors} and commonly denoted as $u$ (\emph{up}), $d$ (\emph{down}) and $s$ (\emph{strange}). They differ from each other in terms of mass and quantum numbers. This aspect is illustrated in Table~\ref{table}. In this picture, baryons (mesons) are seen as the composite states of three quarks (one quark and one antiquark). Modern quark models now include more quark flavors forming heavier hadrons.  

\begin{table}[b]
\caption{$u,d$ and $s$ label the up, down and strange quark. $B$ is the baryon number, $Q$ is the charge, $I_3$ stands for the projection of isospin on an arbitrary axis, while $S_{\! t}$ is the strangeness.}
\centering
\begin{tabular}{c c c c c c l}
\hline
\hline
Quark & Spin & B & Q & $I_3$ & S$_{\! t}$\\
\hline
\hline
${u}$ &  ${1}/{2}$  & ${1}/{3}$ & $+{2}/{3}$  &  $+{1}/{2}$ &   $0$   \\
${d}$ &  ${1}/{2}$  & ${1}/{3}$ & $-{1}/{3}$ & $-{1}/{2}$ &   $0$   \\
${s}$ &  ${1}/{2}$  & ${1}/{3}$ & $-{1}/{3}$ &      $0$        &  $-1$    \\
\hline
\hline
\end{tabular}
	\label{table}
\end{table}

As a first step in the simulation of subatomic particles, here we are interested in nucleons (protons and neutrons), which are the lightest particles in the spectrum of baryons. The proton is a $uud$ bound quark state belonging to the symmetric spin$-\tfrac{1}{2}$ octet (part of the Eightfold way). The neutron also belongs to this octet, but its quark content is  $udd$. From Table \ref{table} one can easily check that these quark contents lead directly to the fact that nucleons are indeed baryons with baryonic quantum number ${B=\tfrac{1}{3}+\tfrac{1}{3}+\tfrac{1}{3}=1}$ and that the proton is charged with ${Q=\tfrac{2}{3}+\tfrac{2}{3}-\tfrac{1}{3}=1}$, while the neutron is neutral. The derivation of the flavor-spin bound quark states involves the standard addition of angular momenta, as described in detail in Ref.~\cite{qlbook}. One is eventually led to the following states describing a spin-up proton or neutron  
\begin{equation}
\label{protonneutron}
\begin{aligned}
|\mathrm{proton}\rangle &=(|p_S\rangle|\chi_S\rangle+|p_A\rangle|\chi_A\rangle)/{\sqrt 2},\\
|\mathrm{neutron}\rangle &=(|n_S\rangle|\chi_S\rangle+|n_A\rangle|\chi_A\rangle)/{\sqrt 2},
\end{aligned}
\end{equation}
where we have introduced the flavor states
$|p_A\rangle=({|udu\rangle{-}|duu\rangle})/{\sqrt{2}},
|p_S\rangle=({|udu\rangle{+}|duu\rangle-2|uud\rangle})/{\sqrt{6}},
$
$|n_A\rangle=({|dud\rangle{-}|udd\rangle})/{\sqrt{2}},
|n_S\rangle=({|dud\rangle{+}|udd\rangle-2|ddu\rangle})/{\sqrt{6}}
$
and the spin states
\begin{equation}
\label{inline}
|\chi_A\rangle{=}({|\!\uparrow\downarrow\uparrow\rangle-|\!\downarrow\uparrow\uparrow\rangle})/{\sqrt{2},}
~|\chi_S\rangle{=}({|\!\uparrow\downarrow\uparrow\rangle+|\!\downarrow\uparrow\uparrow\rangle-2|\!\uparrow\uparrow\downarrow\rangle})/{\sqrt{6}}.
\end{equation}
The subscripts $S$ and $A$ in the states above draw the attention to their symmetry and antisymmetry properties with respect to interchange of the first two quarks. The states in Eq.~(\ref{inline}) are evidently non-separable, thus showing the entanglement in the nucleonic state. However, it is important to remark that such entanglement can not be used directly as a resource given the impossibility to coherently manipulate individual quarks in the nucleons. For symmetric ground states of baryons, a problem arises due to their fermionic nature which enforces their description by overall antisymmetric states. This {led} to the discovery of an additional degree of freedom called color (red, green, or blue), and to the postulate that all hadrons are colorless, i.e. the color degree of freedom is described by an antisymmetric state (color singlet). The color structure of all baryons is thus identical and so we omit it in the following. However, it is important to emphasize that quarks interact strongly by exchanging color. 

\emph{Quantum Circuits.-} 
Let us employ the logical encoding $|u\rangle{\equiv}|0\rangle,|d\rangle{\equiv}|1\rangle, |{\uparrow}\rangle{\equiv}|0\rangle$ and $|{\downarrow}\rangle{\equiv}|1\rangle$. With this notation
\begin{equation}
\label{basic}
\begin{aligned}
|p_A\rangle &=\left(\otimes^3_{j=1}\hat\sigma_{x,j}\right)|n_A\rangle=|\chi_A\rangle=(|010\rangle-|100\rangle)/{\sqrt 2},\\
|p_S\rangle &=\left(\otimes^3_{j=1}\hat\sigma_{x,j}\right)|n_S\rangle=|\chi_S\rangle=(|010\rangle+|100\rangle-2|001\rangle)/{\sqrt 6}.
\end{aligned}
\end{equation}

\begin{figure}[t]
{\bf (a)}\hskip4cm{\bf (b)}\\
\includegraphics[width=3.5cm]{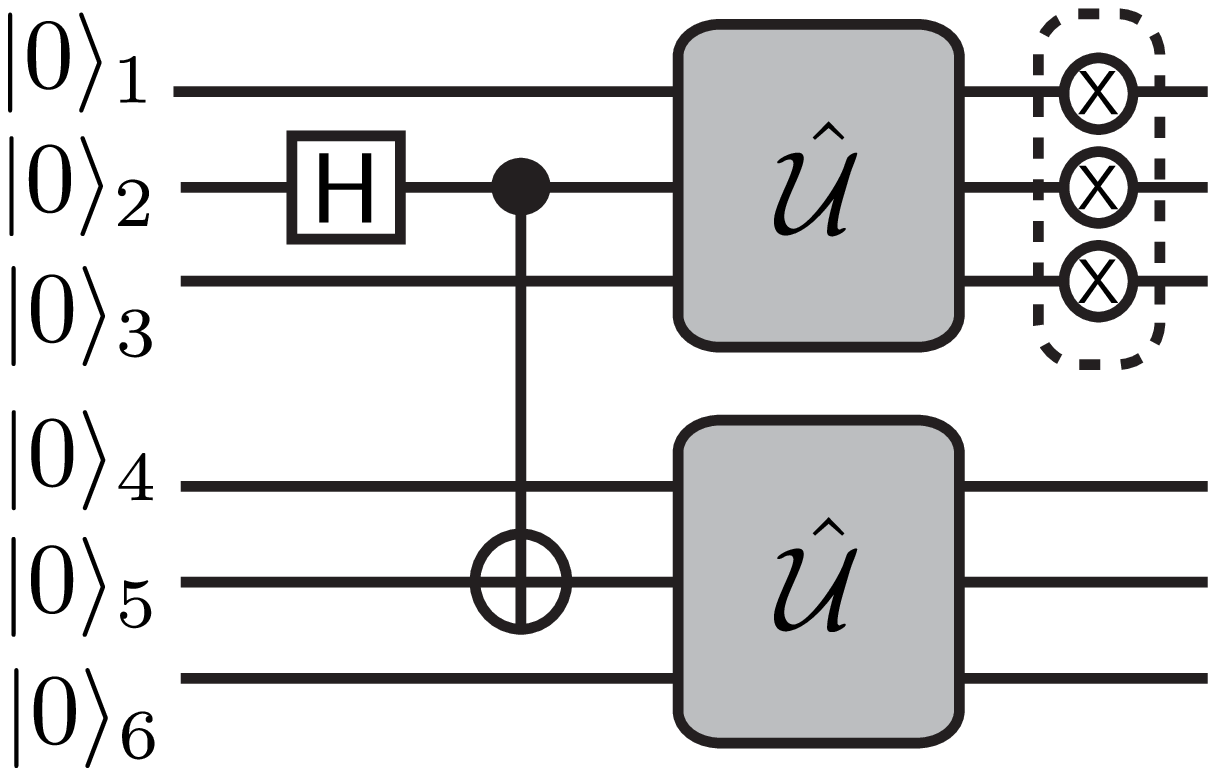}~~~\includegraphics[width=3.5cm]{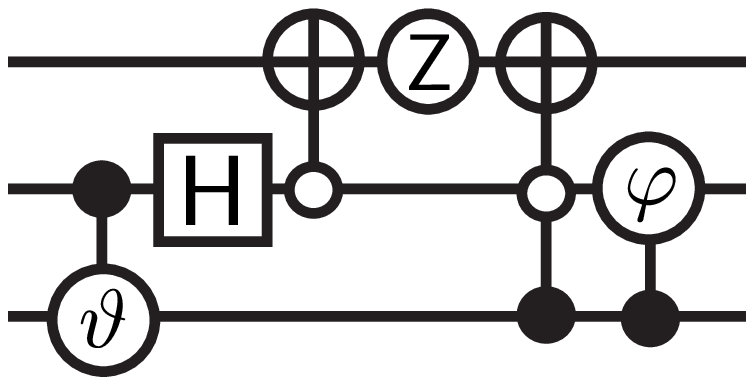}
\caption{{(a)} Circuit for the preparation of $\ket{\text{proton}}$.  We show the symbol for ${\sf H}_2$ and ${\sf CNOT}\!_{(2)5}$. The unitary transformation $\hat{\cal U}$ is decomposed as in {(b)}. We show the symbols for single-qubit phase-shift ${\sf Z}{\equiv}\hat{\sigma}_{z}$, controlled-rotations by angles $\vartheta,\varphi$ and a {\sf CCNOT}. Here, $\vartheta{=}\arccos(-\sqrt{2/3})$ and $\varphi{=}\pi/4$. Empty (filled) dots indicate control operated by state $\ket{0}$ ($\ket{1}$) of the corresponding qubit. The three ${\sf X}\equiv{\hat \sigma}_x$ gates in the dashed box of panel {(a)} are required for the generation of a neutron state [see Eq.~(\ref{protonneutron})].}
\label{scheme}
\end{figure}

The neutron state can be easily obtained from the equations above simply by applying $\sigma_x$ operators to the qubits embodying the flavor degrees of freedom. A detailed construction of the proton state from a fiducial initial state is described in the following paragraphs. In what follows, ${\sf CG}_{(j_1j_2..)k}$ is used to indicate a generic many-qubit controlled gate ${\sf G}$ where $(j_1j_2..)$ are the control qubits determining the action of the gate on the target $k$. The notation $\overline{j}_1$ is used when $\ket{0}_{j_1}$ activates a controlled gate, while $\hat{\sigma}_{\alpha,k}$ indicates the $\alpha$-Pauli matrix ($\alpha{=}x,y,z$) for qubit $k$. For easiness of discussion, it is convenient to refer to the constituents of~(\ref{protonneutron}) as qubits $j{=}1,..,6$. Without affecting the generality of our study, we consider the fiducial state $|\psi_0\rangle=|000\rangle_{123}|000\rangle_{456}$ (other choices would result in minor modifications to what follows). It should be noticed that such a choice is unrelated to the physical quark model for nucleons, but is just a convenient choice for our protocol. We now apply a Hadamard gate ${\sf H}_2{=}(\hat{\sigma}_{z,2}{+}\hat{\sigma}_{x,2})/\sqrt 2$ to qubit 2, followed by a controlled-{\sf NOT} gate ${\sf CNOT}_{\!(2)5}{=}\ket{0}_{2}\bra{0}{\otimes}\hat\openone_5{+}\ket{1}_{2}\bra{1}{\otimes}\hat{\sigma}_{x,5}$ 
 and look for the unitary transformation $\hat{\cal U}$ such that
\begin{equation}
\label{transform}
\begin{aligned}
\ket{p_A}&=\hat{\cal U}\ket{000}_{123},~~\ket{p_S}=\hat{\cal U}\ket{010}_{123},\\
\ket{\chi_A}&=\hat{\cal U}\ket{000}_{456},~~\ket{\chi_S}=\hat{\cal U}\ket{010}_{456}.
\end{aligned}
\end{equation}
In our scheme, the first three (last three) qubits are clearly used to embody flavor (spin). It is worthwhile to notice that the only entangling operation required between flavor and spin qubits is the initial  ${\sf CNOT}_{\!(2)5}$ gate.  There is no need to synthesize a full six-qubit unitary operation, which represents a remarkable simplification in design due, in part, to our choice for information encoding. When the class of allowed gates is restricted to single- and two-qubit ones, any three-qubit unitary operation can be implemented with at most 20 {\sf CNOT} gates and arbitrary single-qubit rotations ~\cite{shende}. In what follows, we provide an explicit quantum circuit that implements the transformations in Eq.~(\ref{transform}) using only 6 single and two-qubit gates. 
A sketch of the whole protocol and the decomposition of $\hat{\cal U}$ are given in Fig.~\ref{scheme}.
Our procedure is as follows: we concentrate on the triplet (1,2,3), although our analysis is applied to (4,5,6) with no changes. The eigenvectors of the reduced density matrix $\varrho_{123}{=}\text{Tr}_{456}(\ketbra{\text{proton}}{\text{proton}})$ all have real components. Therefore, the matrix $\hat{\cal P}$ that diagonalizes $\varrho_{123}$ is special orthogonal and unitary and can embody $\hat{\cal U}$. However, as the only transformations required by our protocol are those listed in Eq.~(\ref{transform}), there is no necessity for reproducing the whole $\hat{\cal P}$: as just $\ket{000}_{123}$ and $\ket{010}_{123}$ are involved, only the first and third columns of $\hat{\cal U}$ have to be identical to the homonymous in $\hat{\cal P}$. This makes the remaining states of the three-qubit computational basis {\it don't-care logical entries} such that $(\hat{\cal U}{-}\hat{\cal P})\ket{q}_{123}{=}0$ [${\neq}0$] for $q{=}0,2$ [$q\neq{0,2}$] (here $q{=}0,..,8$ corresponds to the binary number identifying the elements of the computational basis). 
We thus see that 
\begin{equation}
\label{u}
\hat{\cal U}{\simeq}{\sf CR}_{(3)2}(\varphi){\sf CCNOT}_{\!(3\overline{2})1}{\sf Z}_1{\sf CNOT}_{\!(\overline 2)1}
{\sf H}_2{\sf CR}_{(2)3}(\vartheta),
\end{equation}
where ${\sf CR}_{\!(j)k}(\zeta)$ is a gate rotating the target qubit $k$ by an angle $\zeta$ when the control qubit $j$ is in $\ket{1}_j$. Here, $(\vartheta,\varphi)=(\arccos(-\sqrt{2/3}),\pi/4)$. The $\simeq$ sign in Eq.~(\ref{u}) is due to the fact that the quantum circuit in Fig.~\ref{scheme}{(b)} transforms $\ket{010}_{123}$ ($\ket{010}_{456}$) into $-\ket{p_{S}}$ ($-\ket{\chi_{S}}$). However, this is immaterial to us as $\ket{p_S}\ket{\chi_S}$ appears in the protonic state. We stress the key role played by ${\sf CR}_{\!(2)3}(\vartheta)$: for state $\ket{000}_{123}$ ($\ket{000}_{456}$), this is an inactive gate so that the qubit 3 (6) would never be in $\ket{1}_3$ ($\ket{1}_6$). This makes the sequence ${\sf CR}_{(3)2}(\varphi){\sf CCNOT}_{\!(3\overline{2})1}$ redundant for this entry and $\hat{\cal U}$ becomes a Bell-entangling circuit~\cite{qcb}.

Let us now briefly comment on the resources needed for the simulation of $\hat{\cal U}$. As the rotations used to build up ${\sf CR}_{\!(j)k}(\zeta)$ read $\hat{R}_k(\zeta){=}\sin\zeta\,\hat\sigma_{z,k}{+}\cos\zeta\,\hat\sigma_{x,k}~(\zeta{=}\vartheta,\varphi)$, they can be simulated by means of a single {\sf CNOT} gate and two rotations each as~\cite{barenco}
${\sf CR}_{\!(j)k}(\zeta){=}[\openone_j\otimes\hat{R}_k(\zeta/2)]{\sf CNOT}_{\!(j)k}[\openone_j\otimes\hat{R}_k(\zeta/2)]$.
Moreover, an economic simulation of the Toffoli gate ${\sf CCNOT}_{\!(3\overline{2})1}$ required in our scheme is also possible. Following Ref.~\cite{barenco}, we can simulate a gate congruent to ${\sf CCNOT}_{\!(3\overline 2)1}$ 
as
${{\sf X}_2\hat{W}_1(\frac{\pi}{8}){\sf CNOT}_{\!(2)1}\hat{W}_1(\frac{\pi}{8}){\sf CNOT}_{\!(3)1}\hat{W}^\dag_1(\frac{\pi}{8}){\sf CNOT}_{\!(2)1}\hat{W}^\dag_1(\frac{\pi}{8}){\sf X}_2}$,
where ${\hat{W}_k(\zeta)=\hat{R}_k(\zeta)\hat{\sigma}_{x,k}}$. This gives a truth table identical to the one for ${\sf CCNOT}_{\!(3\overline{2})1}$ but for the entry corresponding to state $\ket{111}$, which are transformed into $-\ket{111}$. However, as this state does not enter into the parts of $\hat{\cal U}$ needed in Eq.~(\ref{transform}), the congruent-gate simulation fulfills our needs. Our decomposition of $\hat{\cal U}$ thus requires only 6 ${\sf CNOT}$'s, far less than the estimated upper bound given in Ref.~\cite{shende}. We believe that, although we cannot claim for optimality and further improvements may be in order, the decomposition we propose could well be seen as rather efficient. The realization of the neutron state goes along the same lines, although it requires three ${\sf X}$ gates on $1,2$ and $3$ [cfr. Fig.~\ref{scheme} {(a)}].

The resource-estimate given above is based on the assumption that only two-qubit interactions would be available in the specific set-up used for the simulation here at hand. However, in many cases this might well be too limiting as genuine multi-qubit interactions could be in order. In NMR, for example, three-body interactions can be efficiently achieved and have been used to investigate ground-state properties at criticality~\cite{tseng-suter}. Analogously, theoretical schemes have been put forward for the achievement of genuine and tunable three-body couplings in triangular-cell optical lattices loaded with two-species cold atoms~\cite{pachos}. This scenario is particularly interesting as it offers the possibility for the combination of laser-induced three-body interactions and measurement-based quantum computing~\cite{briegel} in a lattice of many-body systems. In this context, indeed, a {\it bowtie} lattice-cell configuration has been shown to naturally entail three-body 
 couplings suitable for the economic simulation, via a measurement-based approach, of ${\sf CCNOT}$ gates (along with the standard toolbox of ${\sf CNOT}$ and single-qubit rotations)~\cite{natural}. It is interesting to notice that both the methods at the basis of Refs.~\cite{pachos,natural} rely on the {\it induction} of the couplings $\hat{\cal H}_{\text{c}}{=}\sum_{i}\hat{\sigma}_{x,i-1}\hat{\sigma}_{z,i}\hat{\sigma}_{x,i+1}$, among others, where $i$ is the label for physical qubits in the neighborhood of a given lattice-cell configuration. This is the so-called cluster Hamiltonian~\cite{pachos}, whose ground state encodes a cluster state, which is the key resource for measurement-based processing of information. Therefore, a very promising setting for the simulation of  the nucleonic spin states involves the natural encoding of cluster states in 
lattices of quantum many-body systems coupled by suitable two- or three-particle interactions, as discussed in~\cite{bartlett}, having a lattice-cell structure suitable for the simulation of Toffoli gates. This would allow for the implementation of our proposal with just 13 two-qubit gates or, significantly, only 9 two- and three-qubit gates.  

\begin{figure}[t]
{{\bf (a)}}\hskip4.5cm{{\bf (b)}}\\
\includegraphics[width=4.0cm]{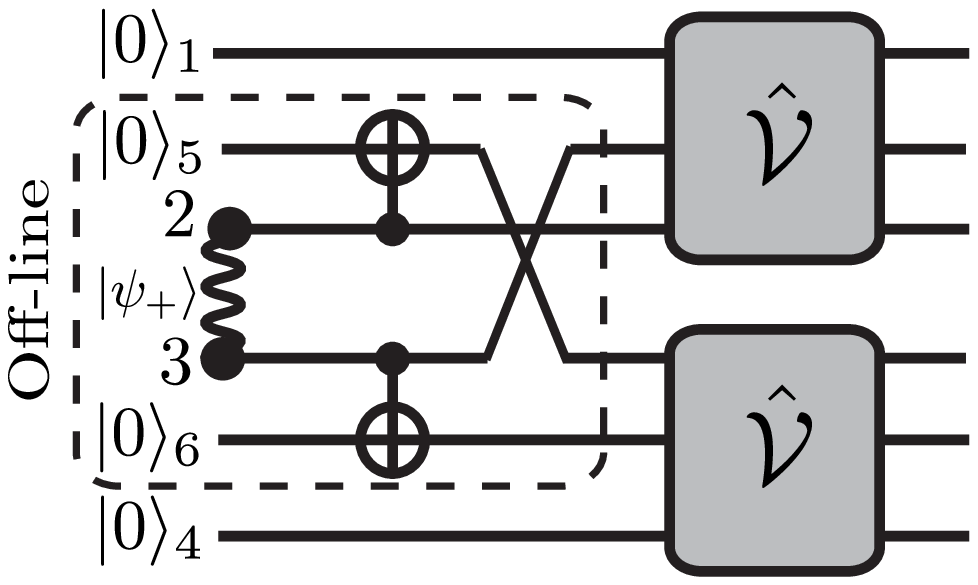}~~\includegraphics[width=4.0cm]{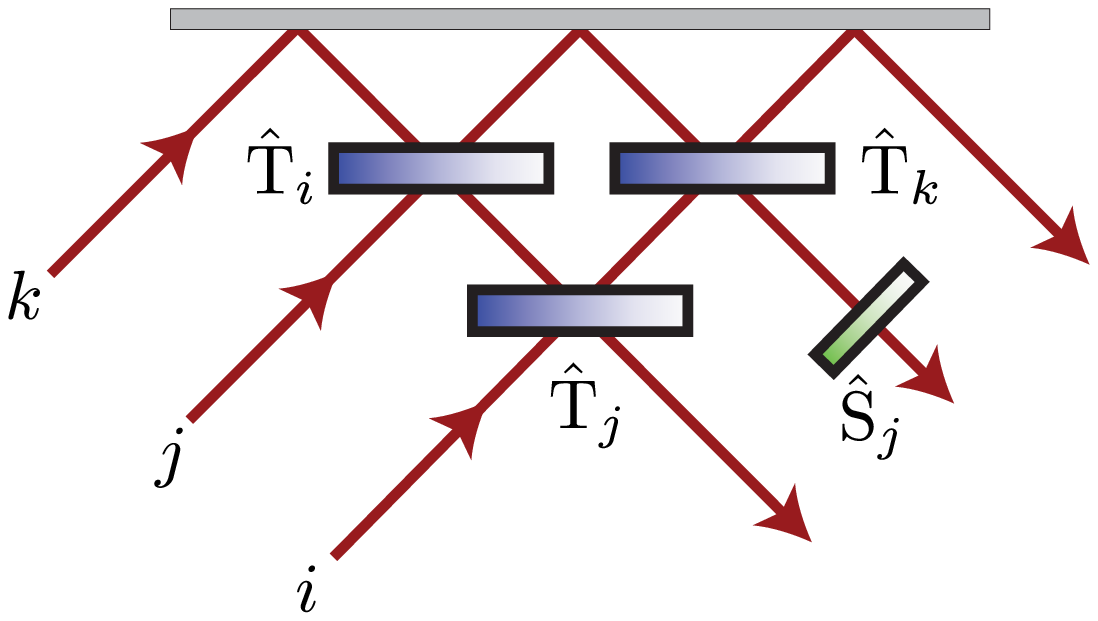}
\caption{{(a)} Protocol for the simulation of a protonic state in an optical network. The single-photon entangled state $\ket{\psi_+}_{23}$ is generated off-line and entangled to modes $5$ and $6$, both prepared in their vacuum state. Transformation $\hat{\cal V}$ completes the protocol. It is realized using three beam splitters and a phase shifter, as shown in panel {(b)} for the mode-triplet $(i,j,k)$ with $i{=}1,4$, $j{=}2,5$ and $k{=}3,6$.}
\label{decoOpt}
\end{figure}

The discussion above does not exhaust all the possibilities for the realization of the nucleonic spin states. In fact, one can easily work out a different version of our protocol that is very suitable for an all-optical implementation. Our starting point is the fact that Eqs.~(\ref{basic}) involve only three elements of the computational basis. The unitary matrix $\hat{\cal U}$ (an $8\times 8$ matrix) in Eq.~(\ref{transform}) can thus be replaced by the more compact unitary (in the ordered basis $\{\ket{001},\ket{010},\ket{100}\}$)
\begin{equation}
\label{compact}
{\hat{\cal V}=
\begin{pmatrix}
0&-\sqrt{2/3}&1/\sqrt 3\\
1/\sqrt 2& 1/\sqrt 6& 1/\sqrt 3\\
-1/\sqrt 2&1/\sqrt 6&1/\sqrt 3
\end{pmatrix}},
\end{equation}
where the first two columns are the components of states $\ket{r_{A,S}}$ ($r{=}p,\chi$) and the last one is determined by imposing unitarity. Eq.~(\ref{compact}) helps in constructing $\ket{\text{proton}}$ if, instead of state $\ket{\psi_1}$, we use $\ket{\psi_2}{=}\ket{00}_{14}(\ket{0101}+\ket{1010})_{2356}/\sqrt{2}$, which can be obtained using $\ket{\psi_+}{=}(\ket{01}+\ket{10})_{23}/\sqrt{2}$ as shown in Fig.~\ref{decoOpt}{(a)}. 

By encoding the states of the basis used to express $\hat{\cal V}$ into spatial modes of light~\cite{commentoencoding}, we deal with a three-mode unitary operation that can be decomposed in terms of beam splitters, phase shifters and rotators as proven in Ref.~\cite{reck}. By writing the beam splitter operations on spatial modes $m$ and $n$ as $\hat{\text{BS}}_{mn}(\omega){=}\cos\omega\hat\sigma_x{+}\sin\omega\hat\sigma_z$, we find $\hat{\cal V}_{ijk}=\hat{\text S}_j\hat{\text T}_k\hat{\text T}_j\hat{\text T}_i$ with $\hat{\text T}_i{\equiv}\hat{\text{BS}}_{jk}(-\arccos(1/\sqrt 3))$, $\hat{\text T}_j{\equiv}\hat{\text{BS}}_{ik}(-3\pi/4)$, $\hat{\text T}_k{\equiv}\hat{\text{BS}}_{ij}(0)$  and $\hat{\text S}_j$ being a $\pi$-phase shift ($i{=}1,4$, $j{=}2,5$ and $k{=}3,6$). Therefore, the simple interferometric setting shown in Fig.~\ref{decoOpt}{(b)} realizes the desired transformation. On the other hand, state $\ket{\psi_2}$ should be fully within the grasp of current experim
 ental abilities in bulk-optics and can be implemented  starting from a Bell state, as generated from a Type-II parametric down-conversion process, and using the handiness of polarization beam splitters as polarization-path ${\sf CNOT}$ gates~\cite{jlo}. Interesting multi-photon states (of up to six photons) are routinely prepared and manipulated in many optical labs~\cite{various}, putting our proposal fully within the realm of realistic and implementable schemes for photonic quantum state engineering.

As an application of the emulated quark states described above, we now discuss the measurement of the magnetic moments. The net magnetic moment of a nucleon is simply the sum 
of the moments of the three constituent quarks, i.e. $\mu_{\cal N}{=}\sum_{i=1}^3\mu_i\langle{\cal N}|\hat\sigma_{z,i}|{\cal N}\rangle$ with ${\cal N}{=}\{\text{proton, neutron}\}$ and $\mu_i$ the magnetic moment of the quarks. By assuming equal mass for the $d$ and $u$ quarks, one has  ${\mu_u=-2\mu_d}$ ~\cite{qlbook} leading to ${\mu_{\mathrm{neutron}}/\mu_{\mathrm{proton}}=-2/3}$, which is in excellent agreement with 
the value determined experimentally. Using our encoding, the nucleon magnetic moment operator  reads 
$\hat\Xi{=}\sum_{i=1}^3(|1\rangle_i\langle 1|{-}2|0\rangle_i\langle 0|)\otimes\hat\sigma_{z,i+3}$, where the summation is over the constituent quarks. As $\hat\Xi$ only involves local operators of qubits $i$ and $i+3$, the measurement of $\langle\hat\Xi\rangle$ will be possible in both the simulation contexts discussed here.

\emph{Conclusions.-} We have proposed a quantum circuit for mimicking the nucleonic spin states resulting from the combinations of their quark components, as predicted by the Eightfold way. It is important to emphasize that dynamics is not accessible in our protocol, which mainly embodies a kinematic model. However, our proposal represents a first attempt to bring the realm of quantum simulation to the elementary-particle domain, which was so far unexplored, to the best of our knowledge. Clearly, the formulation of a quantum mechanical framework for the simulation of particle physics where also the complex aspects of QCD were included would be a very important achievement due to the intrinsic difficulty to simulate it in a classical computer. This is an important and clearly difficult topic to be investigated from now on. Here, we moved on in a different direction, and presented the simulation of nucleonic spin states as given by a simplified kinematic approach of quarks.  We have presented a simple generation scheme involving manipulations which are currently implemented in different experimental setups.  The most demanding part of quantum circuit decomposition proposed here ($\hat{\cal U}$ implementation) requires only 6 ${\sf CNOT}$'s, far less than the estimated upper bound given in Ref.~\cite{shende} for a general three-qubit gate, and we believe that our proposal of decomposition could well be seen as rather efficient, although a general proof is still lacking. 

Our scheme does not include color-exchange through gluon emission and absorption and still has room for further developments, including future circuit decompositions and physical implementations of other hadron simulations. A possibility is the inclusion of heavier quarks or the simulation of mesons. With the addition of strange quarks, for instance, a new scenario would be opened, where baryons such as $\Sigma^{+,-,0}$ could be simulated. As a visionary goal, it would be very interesting to design a unified simulation framework able to encompass more quark bound states. 

{\it Acknowledgments.-} F.L.S. thanks CNPq for
partial support under Grant No. 303042/2008-7 and also
acknowledges partial support of the Brazilian National
Institutes of Science and Technology of Quantum Information
(INCT-IQ). M.P. thanks EPSRC (E/G004579/1) for support.




\begin{thebibliography}{99}

\bibitem{mott} M. Greiner, {\it et al.}, Nature (London) \textbf{415}, 39 (2002).
%
\bibitem{cc} M. J. Hartmann, F. G. S. L. Brand\~ao, and M. B. Plenio, Laser \& Photon. Rev. {\bf 2}, 527 (2008) and references therein.
%
\bibitem{emol} A. Aspuru-Guzik, {\it et al.}, Science \textbf{309}, 1704 (2005); B. P. Lanyon, {\it et al.}, Nature Chem. \textbf{2}, 106 (2009). 
%
\bibitem{bec} L. J. Garay, {\it et al.} Phys. Rev. Lett. {\bf 85}, 4643 (2000); A. Retzker, {\it et al.}, {\it ibid.} {\bf 101}, 110402 (2008); I. Carusotto, {\it et al.}, New J. Phys. {\bf 10}, 103001 (2008).
%
\bibitem{morigi} D. E. Chang, {\it et al.}, Nature Phys. {\bf 4}, 884 (2008).
%
\bibitem{zitter} J. Y. Vaishnav and C. W. Clark, Phys. Rev. Lett. {\bf 100}, 153002 (2008); J. Schliemann, D. Loss, and R. M. Westervelt, {\it ibid.} {\bf 94}, 206801 (2005).
%
\bibitem{dirac} R. Gerritsma, {\it et al.}, Nature (London) \textbf{463}, 68 (2010).
%
\bibitem{feynman} R. Feynman, Int. J. Theor. Phys. \textbf{21}, 467 (1982).
%
\bibitem{deutsch} http://www.wired.com/science/discoveries/news/2007/02/72734.
%
\bibitem{buluta} I. Buluta and F. Nori, Science \textbf{326}, 108 (2009).
\bibitem{qlbook} F. Halzen and A. D. Martin, \textit{Quarks \& Leptons: an Introductory Course in Modern Particle Physics} (John Wiley \& Sons, 1984); D. Griffiths \textit{Introduction to Elementary Particles} (John Wiley \& Sons, Inc., 1987).
%
\bibitem{old} S. Meshkov, C. A. Levinson,and H. J. Lepkin, Phys. Rev. Lett. \textbf{10}, 361 (1963) ; S. Meshkov, G. A. Snow, and G. B. Yodh, Phys. Rev. Lett. \textbf{12}, 87 (1964); C. V. Sastry and S. P. Misra, Phys. Rev. D \textbf{1}, 166 (1970).
%
\bibitem{morpurgo} G. Morpurgo, Phys. Rev. D \textbf{46}, 4068 (1992).
%
\bibitem{su3} M. Gell-Mann, Phys. Lett. \textbf{8}, 214 (1964); G. Zweig, CERN Reports No. TH-401 and TH-412, 1964 (unpublished).
%
\bibitem{qcb} M. A. Nielsen and I. L. Chuang, \textit{Quantum Computation and
Quantum Information} (Cambridge University Press, 2000).
%
\bibitem{shende} V. Shende, S. Bullock, and I. Markov, IEEE Transactions on Computer-Aid Design {\bf 25}, 1000 (2006).
%
\bibitem{barenco} A. Barenco, {\it et al.}, Phys. Rev. A {\bf 52}, 3457 (1995).
%
\bibitem{tseng-suter} C. H. Tseng, {\it et al.}, Phys. Rev. A {\bf 61}, 012302 (1999); X. Peng, {\it et al.}, Phys. Rev. Lett. {\bf 103}, 140501 (2009).
%
\bibitem{pachos} J. K. Pachos and M. B. Plenio, Phys. Rev. Lett. {\bf 93}, 056402 (2004); J. K. Pachos and E. Rico, Phys. Rev. A {\bf 70}, 053620 (2004).
%
\bibitem{briegel} H. J. Briegel, {\it et al.}, Nature Phys. {\bf 5}, 19 (2009).
%
\bibitem{natural} M. S. Tame, {\it et al.}, Phys. Rev. A {\bf 73}, 022309 (2006).
%
%
\bibitem{bartlett} A. C. Doherty and S. D. Bartlett, Phys. Rev. Lett. {\bf 103}, 020506 (2009).
%
\bibitem{commentoencoding} Here $\ket{001}_{ijk}$, $\ket{010}_{ijk}$ and $\ket{100}_{ijk}$  stand for a single photon in the spatial mode $i,j$ and $k$ respectively ($i{=}1,4$, $j{=}2,5$, $k{=}3,6$).
%
\bibitem{reck} M. Reck, {\it et al.}, Phys. Rev. Lett. {\bf 73}, 58 (1994).
%
\bibitem{jlo} P. Kalasuwan, {\it et al.}, arXiv:1003.4291; M. Fiorentino, and F. N. C. Wong, Phys. Rev. Lett. {\bf 93}, 070502 (2004).
%
\bibitem{various} C.-Y. Lu, {\it et al.}, Nature Phys. {\bf 3}, 91 (2007); W. Wieczorek, {\it et al.}, Phys. Rev. Lett. {\bf 103}, 020504 (2009); R. Prevedel, {\it et al.}, {\it ibid.} {\bf 103}, 020503 (2009); M. Radmark, {\it et al.}, {\it ibid.} {\bf 103}, 150501 (2009); G. Vallone, {\it et al.}, Phys. Rev. A (to appear, 2010).
%
\end{thebibliography}
\end{document}